# Monte Carlo simulations with indefinite and complex-valued measures

T D Kieu and C J Griffin,
School of Physics,
University of Melbourne,
Parkville VIC 3052,
AUSTRALIA

## Abstract

A method is presented to tackle the sign problem in the simulations of systems having indefinite or complex-valued measures. In general, this new approach is shown to yield statistical errors smaller than the crude Monte Carlo using absolute values of the original measures. Exactly solvable, one-dimensional Ising models with complex temperature and complex activity illustrate the considerable improvements and the workability of the new method even when the crude one fails.

# Introduction

Numerical simulations have opened up new directions in the study of many interesting problems, offering an alternative and complementary method when the problems are analytically managable and the only systematic method in more complicated problems. However, in the case of general dynamical systems even numerical methods have difficulty due to the fundamental 'sign problem' when the measures of generating functions are not positive definite or are complex [1], invalidating the probabilistic interpretation of conventional Monte Carlo (MC) simulations. Many interesting and important physical systems, unfortunately, belong to this class with the sign problem. Examples include: real-time path integrals of quantum mechanics and quantum field theory, lattice QCD at finite temperature and density, chiral gauge theory, quantum statistical systems with fermions. We will consider later the Ising models in a complex magnetic field or with complex temperature.

Many approaches have been proposed for the sign problem but so far none is satisfactory. Complex Langevin simulations [2] cannot be shown to converge to the desire distributions and often fail to do so. Others [3] are either restricted to too small a lattice, too complicated, or not general enough or speculative.

Following is the crude approach of the average sign [4], upon which we want to improve in the next section.

If the measure $\rho(x)$ of a generating function suffers from sign fluctuation then another positive definite function $\tilde{\rho}(x)$ must be chosen for the MC evaluation of the expectation value of an obsevable $\Theta$,

$$\begin{aligned}\langle\Theta\rangle &= \int_x (\Theta(x)\rho(x)/\tilde{\rho}(x))\tilde{\rho}(x) \Big/ \int_x (\rho(x)/\tilde{\rho}(x))\tilde{\rho}(x), \\ &\equiv \langle\langle\Theta\rangle\rangle/\langle\langle 1\rangle\rangle.\end{aligned} \quad (1)$$

In general it is desirable to choose $\tilde{\rho}$ independently of $\Theta$, and we will concentrate on the estimate of the denominator $\langle\langle 1\rangle\rangle$ because of its appearance in all the measurements. It is a simple variational problem to show that the MC probability density, properly normalised, which minimises the variance of $\langle\langle 1\rangle\rangle$ which is $\sigma/\sqrt{\#\text{independent configurations}}$, where

$$\sigma^2 = \int |\rho(x)/\tilde{\rho}(x) - \langle\langle 1\rangle\rangle|^2 \, \tilde{\rho}(x), \quad (2)$$

must be

$$\tilde{\rho}(x) = |\rho(x)| \Big/ \int |\rho(x)|. \quad (3)$$

Thus the sign of $\rho(x)$ is now treated as part of the quantity whose expectation is to be measured.

However, when the denominator $\langle\langle 1\rangle\rangle$ is vanishingly small, $\sigma^2$, though minimised, is $\approx 1 \gg \langle\langle 1\rangle\rangle$ since $\rho/|\rho| = \pm 1$. Then the evaluation of $\langle\Theta\rangle$ becomes unreliable unless the number of independent configurations is many orders of magnitude greater than the large number $1/\langle\langle 1\rangle\rangle^2$. The fluctuation of sign of the measure over configuration space thus renders ineffective the sampling guided by this crude MC method (3). This is the content of the sign problem.

# A new approach

To deal with complex integrands, of which indefinite measures are special cases, we adopt the definition (2) of the variance extended to the absolute values of complex numbers. Statistical analysis from this definition is the same as the standard analysis; except that the range of uncertainty should now be depicted as the radius of an 'uncertainty circle' centred on some central value in the complex plane. The variational proof leading to (3) also remains intact.

We can write the generating function as integrals over two configurational subspaces $\int_x \rho(x) = \int_X \int_Y \rho(X,Y)$. For example, $X$ and $Y$ are the field values on two non-overlapping sublattices.

We will choose the subspace partition in such a way that the multi-dimensional integral over $Y$ can be evaluated analytically or well approximated:

$$\varrho(X) = \int_Y \rho(X,Y). \tag{4}$$

In the example of the next section, due to the short range interactions, $Y$ is chosen to be the subspace of configurations of *non-interacting* spins residing on even sites of the lattice, and an explicit expression for $\varrho$ can thus be obtained.

As in the last section, one can easily prove that the MC weight $\tilde{\varrho}$ that minimises

$$\sigma'^2 = \int_X |\varrho(X)/\tilde{\varrho}(X) - \langle\langle 1 \rangle\rangle|^2 \, \tilde{\varrho}(X), \tag{5}$$

is

$$\tilde{\varrho}(X) = |\varrho(X)| \Big/ \int_X |\varrho(X)|. \tag{6}$$

It then follows that the variance for this new weight is not bigger than that for $\rho(X,Y)$.

$$\begin{aligned}
\sigma'^2 - \sigma^2 &= \int_X |\varrho(X)|^2 / \tilde{\varrho}(X) - \int_X \int_Y |\rho(X,Y)|^2 / \tilde{\rho}(X,Y), \\
&= \left(\int_X |\varrho(X)|\right)^2 - \left(\int_X \int_Y |\rho(X,Y)|\right)^2, \\
&= \left(\int_X \left|\int_Y \rho(X,Y)\right|\right)^2 - \left(\int_X \int_Y |\rho(X,Y)|\right)^2,
\end{aligned} \tag{7}$$

the second line follows from (3) and (6); the last line is from (4) and always less than or equal to zero because of the triangle inequality. The equality occurs if and only if $\rho(X,Y)$ is semi-definite (either positive or negative) for all $X$-configuration. In particular, when there is no sign problem in the first place, expression (5) yields the same statistical deviation as the crude one.

Our approach is now clear. The measures are first summed over a certain subspace, the integration (4) above, to facilitate some partial phase cancellation. Absolute values of these sums are then employed as the MC sampling weights (6). The numerator in (1) can be obtained from an appropriate derivative of the generating function after the partial summation.

The choice for splitting of the integration domain is arbitrary and its effectiveness depends on the physics of the problem. The better the phase cancellation in the partial sum is, the more important the new MC sampling. One of the most convenient choices is the subspace of configurations over some maximal sublattice so that the partial sum can be exactly evaluated or well approximated. In particular, if the interactions are short-range (not necessary nearest-neighbour), maximal, non-interacting sublattices can always be chosen to provide a natural splitting.

# Illustrative examples

To illustrate our method, we study the one-dimensional Ising model with complex activity or complex temperature. The model is simple, exactly soluble and yet sufficient for our purposes to show the improvements over the crude method and the workability of the new approach when the crude one fails.

Although we are not interested in the models *per se*, they are of much physical relevance. Information about the phase transitions for physical values of parameters in the thermodynamic limits can be learned from the finite-volume partition functions in the complex plane. The Yang-Lee edge singularity [5], the distribution of Fisher zeros of the partition function, and hence their analyticity, in the complex temperature plane [6] have been studied. Furthermore, the 1-D models of complex temperature can also be given the physical interpretation of a two-state quantum tunneling system in real time [7].

Owing to the nearest-neighbour interactions, the lattice can be partitioned into odd and even sublattices, of which the Ising spins $s_i$ $(= \pm 1)$ on site $i (\in$ the sublattice) do not interact with each other. Absolute values of sums of the complex-valued weights over the even sublattice, say, are the new Monte Carlo weights.

In our simulations, periodic boundary condition is imposed on the chains of 128 Ising spins which become 64 spins after the partial summation. Ensemble averages are taken over 1000 configurations, out of $2^{128}$ possible configurations. They are separated by 30 heat bath sweeps which is sufficient for thermalisation and decorrelation in all cases except perhaps one, as will be demonstrated shortly. A heat bath sweep is defined to be one run over the chain, covering each spin in turn. The numbers of trials per sweep are different before and after the partial summation because the numbers of spins to be updated are not the same. Both hot and cold initial configurations are used and will be indicated when needed.

The results for real coupling $J$ and purely imaginary magnetic field $H$ are summarised in Table 1. They are evaluated with the crude probability density

$$\tilde{\rho}(\{s\}) \sim \prod_{\text{all sites}} \exp(Js_i s_{i+1}) \qquad (8)$$

Table 1: Purely imaginary magnetic field. The absolute statistical deviations are in square brackets.

| Coupling | Magnetic Field | | Denominator | Relative deviation |
|---|---|---|---|---|
| 0.1 | $0.1i$ | improved | (0.6242,0.0010) [0.0247] | 3.96% |
| | | crude | (0.4330,0.0295) [0.0285] | 6.57% |
| 0.01 | $0.1i$ | improved | (0.7187,0.0241) [0.020] | 3.06% |
| | | crude | (0.5088,0.0125) [0.0272] | 5.35% |
| 0.01 | $0.3i$ | improved | (0.0237,0.0120) [0.0316] | 118.81% |
| | | crude | (-0.0071,0.0163) [0.0316] | 178.04% |

Table 2: Complex coupling and no external field.

| Coupling | | Denominator | Percentage error |
|---|---|---|---|
| (0.5,0.1) | improved | (0.7720,-0.3107) [0.0175] | 2.11% |
| | crude | (0.5719,-0.2231) [0.0250] | 4.07% |
| (0.1,0.1) | improved | (0.2804,0.9464) [0.0051] | 0.51% |
| | crude | (0.1387,0.4787) [0.0274] | 5.50% |
| (0.01,0.1) | improved | (0.9916,0.1284) [0.0005] | 0.05% |
| | crude | (0.5154,0.0533) [0.0270] | 5.22% |
| (0.01,0.5) | improved | (0.7620,0.6361) [0.0039] | 0.39% |
| | crude | (-0.0155,0.0129) [0.0316] | 156.65% |
| | | (-0.0292,0.0323) [0.0316] | 72.65% |

and also with the improved density

$$\tilde{\varrho}(\{s\}) \sim \prod_{\text{odd sites}} |\cosh(J(s_i + s_{i+1}) + H)|. \tag{9}$$

The denominator measured is proportional to the partition function with different proportionality constants for different MC weights. Thus the relevant quantity here is the relative deviation defined as the ratio of the deviations in square brackets by the magnitudes of central values. Autocorrelation of the denominator as a function of number of sweeps is shown in Fig. 1 for both simulations at particular parameter values.

Table 2 contains the results for complex coupling/temperature with no external field. The improved MC weights are as in (9) but with $H = 0$ and complex $J$; and the crude weights as in (8) with $J$ replaced by its real part.

The use of table look up, which can be employed in discrete spin models, helps reducing the computing time of more complicated improved weights to approximately the same amount of time in dealing with crude weights. The improved MC offers consistently smaller relative deviations. Apart form this, further gains are obtained over the crude method: only half of the spins, those residing on the odd sites of the original lattice, now needed to be considered; less number of sweeps, whose computing time again depends on the number of active spins, required for thermalisation and decorrelation; and acceptance rates are also slightly lower.

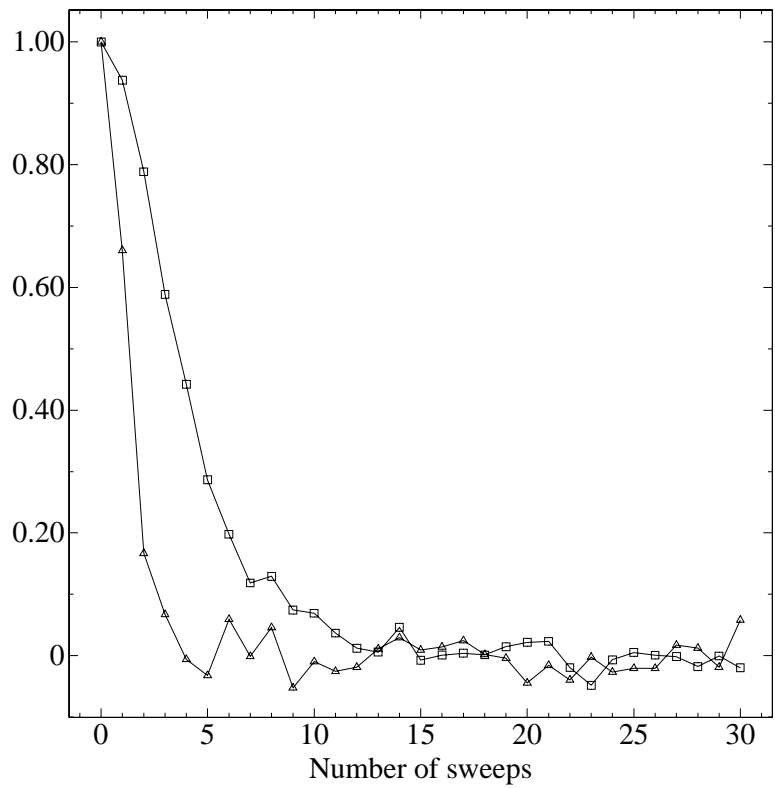

Figure 1: Real parts of the autocorrelation of $\langle\langle 1 \rangle\rangle$ for $J = 0.5$ and $H = 0.1i$ from a total of $2 \times 10^6$ configurations. The boxes are from crude weights; triangles, improved weights.

If the gains in the example of purely imaginary field are about one order of magnitude or less (typically a factor of 6), they are much more significant in the complex temperature illustration.

When the real part of the complex temperature is reduced relative to its imaginary part, the crude MC behaves worse as expected because of the increase in fluctuation of the sign. The gain in relative deviations of the improved over the crude weights is 100 times when $J = (0.01, 0.1)$, for example. This translates into a factor of $10^4$ in configuration number if the error is inversely proprotional to $\sqrt{\#\text{configurations}}$. At the coupling value $J = (0.01, 0.5)$, in particular, the crude MC for both cold and hot starts behaves so badly. But the improved MC continues to work very well. It gives a definite non-zero value; while within the statistically errors given by the former, this value of $J$ could have been taken as a zero of the partition function. The autocorrelation in Fig. 2 shows that the noise is too overwhelming in the crude simulation to tell whether 30 sweeps are sufficient for thermalisation or not.

Of all the coupling values presented in Table 2, the last one also corresponds to the smallest value of the real part of the inverse of correlation length, $\text{Re}(\xi^{-1}) = 0.6$, which is evaluated from the two eigenvalues of the transfer matrix [8]. From these eigenvalues we found that the absolute value of the partition function decreases with the couplings in the order presented [9]. And this is also the reason why the sign problem is getting worse for the crude weight.

We have presented some measurements in Table 3 evaluated by exact, improved MC and crude MC methods respectively. Expressions for the magnetisation and susceptibility are obtained from appropriate derivatives of functions of corresponding partition functions. With the improved weights, we have

$$\left\langle\left\langle \sum s \right\rangle\right\rangle \rightarrow \frac{1}{N_c} \sum_{\text{configs}} \left\{ K(\{s\}) \sum_{\text{odd sites}} [s_i + \tanh(J(s_i + s_{i+1}))] \right\} \qquad (10)$$

and

$$\left\langle\left\langle \left(\sum s\right)^2 \right\rangle\right\rangle \rightarrow$$
$$\frac{1}{N_c} \sum_{\text{configs}} \left\{ K(\{s\}) \left[ \left( \sum_{\text{odd sites}} [s_i + \tanh(J(s_i + s_{i+1}))] \right)^2 \right.\right.$$
$$\left.\left. + \sum_{\text{odd sites}} \frac{1}{\cosh^2(J(s_i + s_{i+1}))} \right] \right\}, \qquad (11)$$

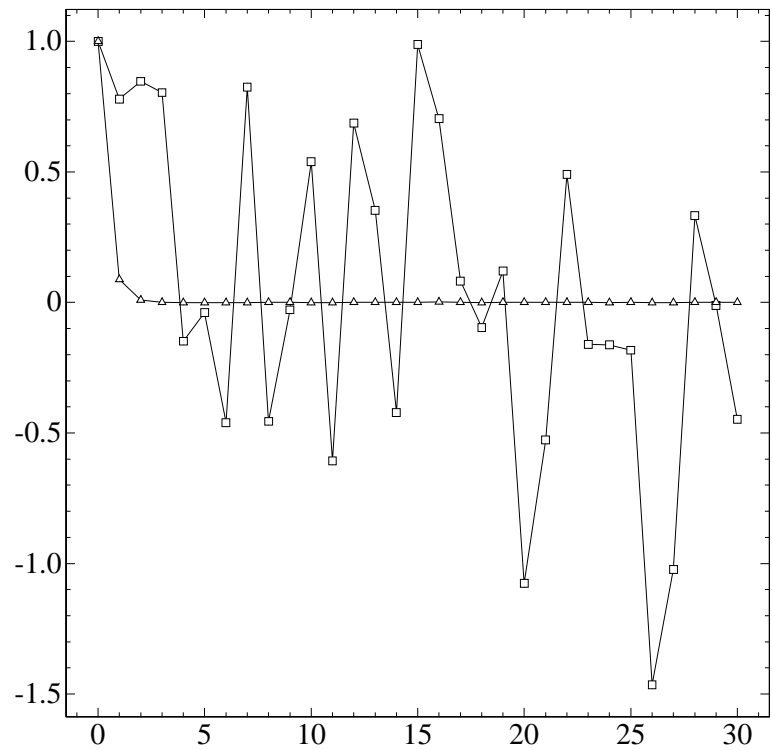

Figure 2: As with Figure 1 but with $J = (0.01, 0.5)$ and $H = 0$.

Table 3: Complex coupling.

| Coupling | | Magnetisation per spin | Susceptibility per spin |
|---|---|---|---|
| (0.1,0.1) | exact | (0,0) | (1.1971,0.2427) |
| | improved | (0.0007,0.0000) [0.0024] | (1.1921,0.2459) [0.0356] |
| | crude | (-0.0032,0.0016) [0.0062] | (1.1651,0.2883) [0.1460] |
| (0.01,0.1) | exact | (0,0) | (0.9999,0.2027) |
| | improved | (-0.0006,-0.0001) [0.0021] | (1.0481,0.2231) [0.0268] |
| | crude | (0.0023,-0.0080) [0.0055] | (1.0868,0.1374) [0.1198] |
| (0.01,0.5) | exact | (0,0) | (0.5512,0.8585) |
| | improved | (-0.0031,-0.0047) [0.0027] | (0.5604,0.8163) [0.0428] |
| | crude | (-0.0522,0.0820) [0.2074] | (0.6808,5.4507) [9.2788] |
| | | (0.0123,-0.0723) [0.0830] | (1.5323,1.3885) [2.3122] |

where $N_c$ is the number of configurations and

$$K(\{s\}) = \prod_{\text{odd sites}} \frac{\cosh(J(s_i + s_{i+1}))}{|\cosh(J(s_i + s_{i+1}))|}. \tag{12}$$

We have also numerically summed one more spin on the remaining odd sublattice to further improve the improved MC weight, but there is no significant gain over the results presented above.

# Concluding remarks

We have presented a method towards a solution for the sign problem. Owing to the sign cancellation in the partial sums, our approach can offer substantial improvements over the crude average sign method and may work even when the later fails, in the region of long correlation length and vanishing partition function. A particular splitting for the summation is chosen for our illustrative examples. And this is the natural choice of splitting which always exists for short-ranged interactions. But other choices of splitting are feasible and how effective they are depends on the physics of the problems. The approach of Ref. [7] could be considered as a special case with a particular and non-trivial splitting where the inner integration, over $Y$ in our notation, was approximated in a certain manner.

When the quantity to be averaged is not smooth on the length scale of the crude weight function, there is an additional source of systematic error in the average sign method. The cancellation in the partial sums may reduce this error by reducing the difference in length scales of the measured quantities and that of the sampling weights.

Lastly on a speculative note, the partial sums might also be used as some pre-conditioning for the complex Langevin simulation which does not converge to the raw complex measures.

# Acknowledgements

We wish to thank Chris Hamer, Bruce McKellar, Mark Novotny and Brian Pendleton for discussions and support. TDK acknowledges the support of a Research Fellowship from the Australian Research Council.